# Cascadable all-optical NAND gates using diffractive networks

*Yi Luo[1,2,3]*    e-mail: yluo2016@ucla.edu

*Deniz Mengu[1,2,3]*    e-mail: denizmengu@g.ucla.edu

*Aydogan Ozcan[1,2,3]\**    e-mail: ozcan@ucla.edu

[1]Electrical and Computer Engineering Department, University of California, Los Angeles, California 90095, USA

[2]Bioengineering Department, University of California, Los Angeles, California 90095, USA

[3]California Nano Systems Institute (CNSI), University of California, Los Angeles, California 90095, USA

\*Correspondence: Prof. Aydogan Ozcan

E-mail: ozcan@ucla.edu

**Abstract**

Owing to its potential advantages such as scalability, low latency and power efficiency, optical computing has seen rapid advances over the last decades. A core unit of a potential all-optical processor would be the NAND gate, which can be cascaded to perform an arbitrary logical operation. Here, we present the design and analysis of cascadable all-optical NAND gates using diffractive neural networks. We encoded the logical values at the input and output planes of a diffractive NAND gate using the relative optical power of two spatially-separated apertures. Based on this architecture, we numerically optimized the design of a diffractive neural network composed of 4 passive layers, spread over an axial length of $250\lambda$, to all-optically perform NAND operation using diffraction of light at a wavelength of $\lambda$, and cascaded these diffractive NAND gates to perform complex logical functions by successively feeding the output of one diffractive NAND gate into another. We demonstrated the cascadability of our diffractive NAND gates by using identical diffractive designs to all-optically perform AND and OR operations, as well as a half-adder. Cascadable all-optical NAND gates composed of spatially-engineered passive diffractive layers can serve as a core component of various optical computing platforms.

# Introduction

Computation forms the backbone of modern society while consuming a large amount of energy. New computation hardware with a smaller carbon footprint and higher computing speed is in great need[1,2]. Through a fruitful history[3,4], optical computing has gained increasing attention in the past decades owing to its potential advantages such as scalability, low latency, and power-efficiency[5,6]. For example, integrated photonic circuits have been designed to accelerate neural network computing in data centers[7–9]. Various optical implementations of task-specific non-von-Neumann computers, such as spiking neural networks[10,11] and reservoir computers[12], have been developed to solve different computational problems. These specialized optical circuitry or networks lit up the future of optical computing, and yet showed relatively limited capability in performing general purpose computation.

An important building block for universal computing is the logical NAND gate, as it can be cascaded to perform any logic operation. Numerous studies have been reported so far on all-optical realizations of NAND gates[13–24]. However, none of these previous optical approaches provided sufficient levels of cascadability to form complex logical circuits using their designed NAND gates as the building block of an optical processor. Some of these earlier techniques relied on nonlinear interferometers[13–16], photonic crystals[17–19], and plasmonic devices[20–22], which brought stringent requirements on the input optical signals in terms of phase, intensity, and/or polarization states of light[6], which partially complicate the hardware design, and more importantly, limit the cascadability of the optical NAND gates. Another set of implementations used micro-ring resonators[23] that encode the input and output logical values into different frequencies, also creating fundamental challenges for cascadability.

Diffractive deep neural networks ($D^2NNs$) are optical computing platforms that use coherent light to process the information encoded in the phase and/or amplitude channels of an input field-of-view[25–30]. The incident light from an input plane (encoding the information) propagates through successive passive diffractive layers, each of which comprises thousands of individual modulation units (termed neurons) that alter the phase and/or amplitude of the light at their corresponding location. The free-space propagation and engineered light-matter interaction through a diffractive network collectively perform a computational task that is learned through a one-time training process. These passive diffractive layers are designed (trained) in a computer using conventional deep learning tools, e.g., stochastic gradient descent and error back-propagation. Once the design converges through this deep learning-based training phase, the diffractive layers are fabricated to form a passive, physical computing unit that does not consume computing power except for the illumination light. Other than performing statistical inference tasks such as object classification[25,28,31,32] and image reconstruction[33,34], enabled by their data-driven training process, diffractive neural networks also show great potential in designing non-intuitive, task-specific, deterministic optical elements, including e.g., spectral filters[35] and pulse shapers[36]. The broad design space provided by diffractive neural networks was also utilized to design logic gates[24]; however, similar to former all-optical implementations discussed earlier[13–23], cascadability was not feasible in this earlier work, limiting its use as a diffractive building block to optically perform an arbitrary logic operation.

In this paper, we present a cascadable all-optical NAND gate based on diffractive neural networks (Fig. 1(a)). We encoded the input/output optical logical values in the relative power of two spatially-separated apertures, where the aperture corresponding to the larger optical signal determines whether the value is 0 (the bottom aperture signal is larger) or 1 (the top aperture signal is larger). A four-layered diffractive network was trained to all-optically perform NAND operation on two optical input logical values (Fig. 1(b)). By projecting the output optical signal/wave of one diffractive NAND gate onto the input aperture of another diffractive NAND gate, one can cascade these NAND gates to perform complex logical computations (Fig. 1(c)). We numerically demonstrated the cascadability of our diffractive NAND gate design by building all-optical AND and OR gates, as well as a half-adder, using identical diffractive NAND gates, one feeding another. The cascadability of the resulting diffractive NAND gate design might serve as a core feature to be able to design various all-optical processors.

## Results
### Diffractive NAND gate design and numerical testing

One method to encode optical logical values of a NAND gate can be to use an additional pump/probe light[15,19,20] in order to successfully handle the case of both input digits being zero, i.e. $\text{NAND}(0,0) = 1$. However, the existence of the probe light creates challenges in the cascadability of a NAND gate while also consuming more energy. Instead, here in this work, we encoded the logical states of our diffractive network into the spatial distribution of the optical power (Fig. 1(a)). Each optical logical value is represented by the power within a pair of encoding apertures placed in a vertical column (Fig 1(a)). The upper/top aperture denotes *True* (T) and the lower one denotes *False* (F) signal, and the larger optical signal determines the logical state. For example, True is encoded with more power in the top aperture when compared with the optical power of the lower aperture. Based on this definition, either one of the logical states (T, F) can inject the same amount of energy into the successive NAND gate that is cascaded. In our design, each encoding aperture was selected to have a size of $4\lambda \times 4\lambda$, and the vertical separation between the two apertures was designed to be $2\lambda$, where $\lambda$ is the wavelength of the incidence light, which can be selected at any part of the electromagnetic spectrum based on the availability of sources and high-resolution fabrication methods. Ideally, an optical logical value has uniform intensity within the corresponding encoding aperture, while no light should be present in the other aperture. *However*, deviations from this ideal scenario in successive, cascaded diffractive NAND gates do not create an issue as the definition of our encoding scheme compares the total power in each encoding aperture regardless of the uniformity of the optical wave intensity.

A four-layered diffractive neural network was trained to perform the NAND operation. The input plane of our diffractive NAND gate had two ports hosting the input optical logical values, where the two columns of apertures were placed side-by-side, with a horizontal separation of 2λ. This 2×2 aperture grid was placed at the center of the input plane, with zero transmittance elsewhere (i.e., blocking regions surrounding the signal apertures). The output port of the diffractive NAND gate was designed at the center of the output field of view of the network. This diffractive NAND gate architecture was iteratively trained using conventional deep learning-based optimization tools in a computer, during which only ideal optical logical values with uniform intensity profiles were used as inputs. A training loss function was applied to the diffractive network's output field-of-view, comparing the resulting optical waves within the output apertures with the ideal output profiles. Further details regarding the training of this diffractive NAND gate can be found in the Methods section.

After its training/design phase, the success of the all-optical NAND computation was first demonstrated by numerically testing it with ideal input optical logical values. The input plane of the diffractive NAND gate with different input values, i.e., $(x_1^0, x_2^0)$ = (T,T), (T,F), (F,T) and (F,F), as well as the corresponding optical intensity profiles at the output plane of the diffractive network (with output logical values of $x_i^1$=(F, T, T, T) respectively) are reported in Fig. 1(b). The notation $x_1^0$ ($x_2^0$) denotes a set of ideal optical logical values, with uniform intensity in each aperture, being injected into the diffractive NAND gate using the left (right) input port at the input plane. Similarly, the notation $x_i^1$ denotes the set of output optical logical values that result from a diffractive NAND gate, taking ideal optical logical values as its inputs. The subscript $i$ indicates that the output optical signals/waves can be projected to either input port (left or right) of another diffractive NAND gate that is cascaded. The superscripts in both the input and output notations denote the 'level' of the optical signals, where the ideal input values are denoted as level 0 and the output optical signals resulting from two ideal input values are denoted as level 1. Therefore, this notation represents a collection of all the possible optical logical values that share the same origin. For example, $x_i^1$ represents all four possible output optical fields generated using combinations of level 0 input optical logical values (($x_1^0, x_2^0$) = (T,T), (T,F), (F,T) and (F,F)).

**Cascading diffractive NAND gates to all-optically perform logic operations**

A unique feature of the presented diffractive NAND gate lies in its cascadability. One NAND gate's output optical wave can be injected into another diffractive NAND gate's input ports for further computation through its diffractive layers. In this context, we would like to further extend the "level" definitions to better describe the diverse origins of the optical signals/waves that are cascaded into successive diffractive NAND gates. If we assume a diffractive NAND gate uses the input optical logical values from levels a and b (i.e., the input optical logical values belong to sets $x_1^a$ and $x_2^b$), then the corresponding output level shall be defined as level (a,b), i.e., the output optical wave lies within the set of $x_i^{a,b}$. Given the fact that a diffractive NAND gate, after its training phase, does not necessarily possess output wave symmetry, $x_i^{a,b}$ and $x_i^{b,a}$ are different sets of optical signals/waves. For example, the output optical logical values calculated from input waves that belong to $x_1^1$ and $x_2^0$ should be denoted as $x_i^{1,0}$, and it is different from the set of output waves defined by $x_i^{0,1}$.

Based on this notation, next we numerically demonstrate that our diffractive NAND gate design can successfully perform logic operations with input values arising from different levels of cascading. Figure 2(a) reports all the possible combinations of optical wave cascading using level 0, level 1, level (1,0), level (0,1) and level (1,1) as inputs to our diffractive NAND gate. For example, at level 0 we have $2^2 = 4$ different optical waves that can result at the output of the diffractive NAND gate using ideal uniform input waves; these four optical waves are then combined into the same set along with the ideal uniform optical inputs, which result in a total of $(4+2)^2 = 36$ unique optical waves at the output of the diffractive NAND. Following the same flow of logic, at the next level of cascading, we have in total $(36+2)^2 = 1444$ different optical waves representing logical values. For these 1444 different wave combinations represented in Fig. 2(a), the calculation/inference accuracy at the output of the diffractive NAND gate is measured to be 91.48%. The black squares in Figure 2(a) indicate the input wave combinations leading to a miscalculation (i.e., an incorrect inference) at the output of the diffractive NAND gate (i.e., ~8.52% of the cases out of 1444 different combinations reported). The output plane intensity profiles of the optical waves that belong to sets $x_i^1$, $x_i^{1,0}$, $x_i^{0,1}$ and $x_i^{1,1}$ and their corresponding input waves are also shown in Fig. 2(b). In fact, further cascading of diffractive NAND gates is also possible: using all the optical logical values and the output optical waves presented in Fig. 2(a), the diffractive NAND gate inference accuracy over the $(1444+2)^2 = 2{,}090{,}916$ different input wave combinations is found to be 80.30%.

These random inference errors that are observed in e.g., Figure 2(a) do *not* constitute a roadblock for synthesizing an all-optical logic processor using cascaded diffractive NAND gates; instead, these error maps (e.g., the black squares in Fig. 2(a)) actually serve as a *design map/guide* for properly cascading diffractive NAND gates to build a logic operator without hitting any one of these error points. Stated differently, by knowing these input-output maps that reveal these rare combinations of faulty diffractive computing points, one can correctly design a diffractive logic processor composed of cascaded all-optical NANDs that avoid using these error points identified in the design guide (Fig. 2(a)). To shed more light on this cascadability design map, we first built basic logical gates (i.e., AND and OR gates) performed by cascading of our diffractive NAND gate. Mathematically, a logical AND operation can be formulated using NAND operations as shown in Fig. 3(a), i.e.,

$$\text{AND}(I_1, I_2) = \text{NAND}\bigl(\text{NAND}(I_1, I_2), \text{NAND}(I_1, I_2)\bigr). \qquad (1)$$

Therefore, an all-optical implementation of the AND operation can be realized with two diffractive NAND gates and a beam splitter (see Fig. 3(b)). $x_1^0$ and $x_2^0$, with uniform optical intensity within the input apertures, were injected into both input ports of the first NAND gate. A beam splitter duplicated the resulting output wave with a 50/50 splitting ratio, and the duplicated logical values were used as the input waves for the second NAND gate that is cascaded. For different input logical value combinations, i.e., $(x_1^0, x_2^0)$ = (T,T), (T,F), (F,T) and (F,F), the correctness of the intermediate all-optical calculation steps $(x_i^1)$ and the output of the AND gate (O) are shown in Fig. 3(c). The intensity profiles at the output plane of the final diffractive NAND gate are also shown in Fig. 3(d), demonstrating the success of the cascaded diffractive NAND system for all-optically performing AND operation.

A similar demonstration of an all-optical OR gate that is composed of cascaded diffractive NAND gates is shown in Fig. 4. OR operation can be formulated using NAND gates as shown in Fig. 4(a), i.e.,

$$\text{OR}(I_1, I_2) = \text{NAND}\big(\text{NAND}(I_1, I_1), \text{NAND}(I_2, I_2)\big). \qquad (2)$$

This all-optical OR gate implementation uses three diffractive NAND gates, as shown in Fig. 4(b). In this case, each input optical logical value was first duplicated using beam splitters and injected into both input ports/apertures of the corresponding diffractive NAND gate. The outputs of the two diffractive NAND gates were then cascaded onto the input ports/apertures of the last diffractive NAND gate, as shown in Fig. 4(b). For different input combinations, the correctness of the intermediate calculation steps $(x_i^1)$ and the output of the OR gate (O) are shown in Fig. 4(c). The intensity profiles at the output plane of the final diffractive NAND gate are also shown in Fig. 4(d), demonstrating the success of the cascaded diffractive system for all-optically performing OR logic operation.

As another basic logic function, NOT operation can be formulated as:

$$\text{NOT}(I) = \text{NAND}(I, I). \tag{3}$$

Therefore, NOT can be realized using a single NAND gate without the need for diffractive cascading. In fact, since it is solely based on the correct implementation of the NAND gate, the accuracy of the NOT calculation using our diffractive NAND design has already been validated through the first and last columns of Fig. 1(b) as well as the *diagonal* elements of the optical logical values $x_i^1$ shown in Fig 2(b).

Other than these basic logical operations (AND, OR, NOT), we also designed an all-optical implementation of a half-adder using cascaded, five diffractive NAND gates. A half-adder adds two binary input values ($I_1$ and $I_2$) and returns their sum (S) and the carry (C). Figure 5(a) presents an electronic realization of a half-adder using 5 NAND gates. The same logical circuit design can be all-optically implemented using cascaded diffractive NANDs, as shown in Fig. 5(b). Based on our design map/guide shown in Fig. 2(a), one can see that a correct calculation of the sum can only be achieved when the NAND gate IV takes the input optical logical values from the set of $x_i^{1,0}$. The correct routing of different optical waves was accordingly optimized to avoid the inference errors marked in Fig. 2(a). A portion of the design map reflecting this optimized optical signal routing is reported in Fig. 5(c), showing the correctness of the intermediate calculation steps. The output intensity profiles of the sum (diffractive NAND gate IV) and the carry (diffractive NAND gate V) are also shown in Fig. 5(d), together with the corresponding inputs, demonstrating the success of the cascaded diffractive system for all-optically performing a half-adder operation.

**Discussion**

In our optical designs and numerical simulations reported so far, the diffractive NAND gates are cascaded to each other using ideal optical projection systems to map an output optical field of a diffractive NAND gate onto another input aperture of a cascaded diffractive NAND gate. However, it is not critical to use high numerical aperture (NA) image projection systems when building complex computational units formed by cascaded diffractive networks. A low NA projection system applies a low pass filter to the output profile of a diffractive gate, which in fact makes it similar to the ideal input signals, with a more uniform intensity profile within each input aperture. To shed more light on this phenomenon, we simulated the cascading of different diffractive NAND gates with projection systems that have lower NA values, i.e., 0.9, 0.75, 0.5 and 0.25. The all-optical calculation accuracy using level 0, level 1, level (1,0), level (0,1) and level (1,1) inputs, i.e., representing 1444 unique combinations of input optical waves, were found to be 91.48%, 91.27%, 90.72%, and 90.86%, respectively. These numerical results demonstrate the robustness of the diffractive NAND gate to a reduction in the NA of the cascading optical projection system between successive diffractive networks.

In summary, we illustrated a cascadable all-optical NAND gate design based on diffractive networks. Successful demonstrations of basic logical gates (AND, OR), as well as a half-adder, were reported using cascaded diffractive NAND gates, which might serve as the building block of next generation optical computing systems.

## Methods
### Forward propagation model

The input plane of a diffractive NAND gate is positioned at $z=0$ and provides a complex optical field $u_0(x, y, z = 0)$ that propagates within the diffractive network. The propagation within a diffractive network is modeled following the Rayleigh-Sommerfeld equation,[25]

$$n_0(x, y, z_0) = u_0(x, y, 0) * w(x, y, z_0), \quad (4)$$

where $n_0$ represents the optical wave right before the first diffractive layer and $*$ denotes 2D convolution operation. $w$ is the complex-valued propagation kernel, which is given by,

$$w(x, y, z) = \frac{z}{r^2}\left(\frac{1}{2\pi r} + \frac{1}{j\lambda}\right)\exp\left(\frac{j2\pi r}{\lambda}\right) \quad (5),$$

with $r = \sqrt{x^2 + y^2 + z^2}$, $j = \sqrt{-1}$ and $\lambda$ being the illumination wavelength. The resulting optical field is further modulated by the spatially-engineered diffractive layers. Assuming each diffractive layer (positioned at $z = z_l, l = 1, \ldots, 4$) to be a thin phase element, the wave modulation provided by the $l^{th}$ diffractive layer can be formulated as:

$$t_l = \exp(j\phi(x, y, z_l)). \quad (6)$$

Therefore, the optical field right after each diffractive layer can be written as

$$u_l(x, y, z_l) = t_l(x, y, z_l) \cdot [u_{l-1}(x, y, z_{l-1}) * w(x, y, z_l - z_{l-1})] \quad (7).$$

After being modulated by all the four diffractive layers of a NAND gate design, the optical field further propagates to the output plane, located at $z = z_d$, which can be written as:

$$o(x,y) = u_4 * w(x, y, z_d - z_4) \ . \tag{8}$$

**Diffractive NAND gate training**

Each diffractive layer of our NAND gate design contains 80×80 neurons (diffractive features) that provide structured phase modulation with a pixel pitch of λ/2. The axial distance between the input plane and the first diffractive layer, between successive diffractive layers, and between the last diffractive layer and the output plane of the NAND is selected to be 50λ. In the network training phase, each batch contained 80 pairs of ideal optical logical values (with uniform intensity profiles) as inputs to the diffractive network, and 200 batches formed one epoch. The logical state of each input value was randomly assigned to be True with ~70% probability and the rest ~30% to be False in order to ensure that the output of a NAND gate has an equal probability of being True or False during the training phase. The optical fields within the designated apertures for the output optical logical values were used to calculate the training loss. For example, if the output value should be *True* (*False*), the upper (lower) aperture should be denoted as $A_{correct}$ and the optical field within the corresponding aperture should be denoted as $o_{correct}(x,y)$, while the other remaining aperture is referred to as $A_{wrong}$, with the corresponding field denoted as $o_{wrong}(x,y)$. Note that the *correct* and *wrong* subscripts in this notation do *not* represent the output logical state but instead indicate if the corresponding aperture label matches the logical calculation result. Based on this notation, our training loss function can be expressed as:

$$Loss = \alpha \times L_{correct} + \beta \times L_{wrong} + \gamma \times L_{unif}. \tag{9}$$

where $L_{correct}$ calculates the $l_2$ distance between the output intensity profile in the *correct* aperture and a plane wave with unit amplitude, i.e.,

$$L_{correct} = \sum_{x,y \in A_{correct}} (1 - |o_{correct}(x,y)|^2)^2 \ . \tag{10}$$

$L_{wrong}$ was accordingly defined to represent the optical power that resides in the *wrong* aperture, i.e.,

$$L_{wrong} = \sum_{x,y \in A_{wrong}} |o_{wrong}(x,y)|^2 . \tag{11}$$

$L_{unif}$, on the other hand, quantified the phase ($\phi$) variation of the optical field within only the *correct* aperture, i.e.,

$$L_{unif} = std\left(\phi(o_{correct}(x,y))\right). \tag{12}$$

The relative weights $\alpha$, $\beta$ and $\gamma$ in Eq. (9) were empirically selected to be 100, 10, and 50, respectively. After calculating the loss values, the phase profiles on each diffractive layer were updated using an Adam optimizer[37], which concludes one training batch. The diffractive NAND gate model was trained using Python (v3.7.3) and TensorFlow (v.1.15.0, Google Inc.) for 50 epochs on a desktop computer, with a GeForce GTX 1080 Ti graphical processing unit (GPU, Nvidia Inc.), an Intel® Core™ i9-7900X central processing unit (CPU, Intel Inc.) and 64 GB of RAM. This training process of a diffractive NAND gate takes ~100 min to complete (50 epochs).

**References**

1   Belkhir L, Elmeligi A. Assessing ICT global emissions footprint: Trends to 2040 & recommendations. *Journal of Cleaner Production* 2018; **177**: 448–463.

2   Liu Y, Wei X, Xiao J, Liu Z, Xu Y, Tian Y. Energy consumption and emission mitigation prediction based on data center traffic and PUE for global data centers. *Global Energy Interconnection* 2020; **3**: 272–282.

3   Athale R, Psaltis D. Optical Computing: Past and Future. *Optics & Photonics News* 2016; **27**: 32.

4   Ambs P. Optical Computing: A 60-Year Adventure. *Advances in Optical Technologies* 2010; **2010**: 1–15.

5   Caulfield HJ, Dolev S. Why future supercomputing requires optics. *Nature Photon* 2010; **4**: 261–263.

6   Miller DAB. Are optical transistors the logical next step? *Nature Photon* 2010; **4**: 3–5.

7   Shen Y, Harris NC, Skirlo S, Prabhu M, Baehr-Jones T, Hochberg M *et al.* Deep learning with coherent nanophotonic circuits. *Nature Photonics* 2017; **11**: 441–446.

8   Hamerly R, Bernstein L, Sludds A, Soljačić M, Englund D. Large-Scale Optical Neural Networks Based on Photoelectric Multiplication. *Phys Rev X* 2019; **9**: 021032.

9   Xu X, Tan M, Corcoran B, Wu J, Boes A, Nguyen TG *et al.* 11 TOPS photonic convolutional accelerator for optical neural networks. *Nature* 2021; **589**: 44–51.

10  Feldmann J, Youngblood N, Wright CD, Bhaskaran H, Pernice WHP. All-optical spiking neurosynaptic networks with self-learning capabilities. *Nature* 2019; **569**: 208–214.

11  Nahmias MA, Peng H-T, de Lima TF, Huang C, Tait AN, Shastri BJ *et al.* A Laser Spiking Neuron in a Photonic Integrated Circuit. *arXiv:201208516 [physics]* 2020.http://arxiv.org/abs/2012.08516 (accessed 22 Dec2020).

12  Teğin U, Yıldırım M, Oğuz İ, Moser C, Psaltis D. Scalable optical learning operator. *Nat Comput Sci* 2021; **1**: 542–549.

13  Hall KL, Rauschenbach KA. 100-Gbit/s bitwise logic. *Opt Lett* 1998; **23**: 1271.

14 Dimitriadou E, Zoiros KE. On the feasibility of ultrafast all-optical NAND gate using single quantum-dot semiconductor optical amplifier-based Mach–Zehnder interferometer. *Optics & Laser Technology* 2012; **44**: 1971–1981.

15 Datta K, Chattopadhyay T, Sengupta I. All optical design of binary adders using semiconductor optical amplifier assisted Mach–Zehnder interferometer. *Microelectronics Journal* 2015; **46**: 839–847.

16 Kim SH, Kim JH, Yu BG, Byun YT, Jeon YM, Lee S *et al.* All-optical NAND gate using cross-gain modulation in semiconductor optical amplifiers. ; : 2.

17 Fu Y, Hu X, Gong Q. Silicon photonic crystal all-optical logic gates. *Physics Letters A* 2013; **377**: 329–333.

18 Alipour-Banaei H, Serajmohammadi S, Mehdizadeh F. All optical NOR and NAND gate based on nonlinear photonic crystal ring resonators. *Optik* 2014; **125**: 5701–5704.

19 Alipour-Banaei H, Serajmohammadi S, Mehdizadeh F. All optical NAND gate based on nonlinear photonic crystal ring resonators. *Optik* 2017; **130**: 1214–1221.

20 Wei H, Wang Z, Tian X, Käll M, Xu H. Cascaded logic gates in nanophotonic plasmon networks. *Nature Communications* 2011; **2**. doi:10.1038/ncomms1388.

21 Wei H, Li Z, Tian X, Wang Z, Cong F, Liu N *et al.* Quantum Dot-Based Local Field Imaging Reveals Plasmon-Based Interferometric Logic in Silver Nanowire Networks. *Nano Letters* 2011; **11**: 471–475.

22 Sang Y, Wu X, Raja SS, Wang C-Y, Li H, Ding Y *et al.* Broadband Multifunctional Plasmonic Logic Gates. *Advanced Optical Materials* 2018; **6**: 1701368.

23 Xu Q, Lipson M. All-optical logic based on silicon micro-ring resonators. *Opt Express, OE* 2007; **15**: 924–929.

24 Qian C, Lin X, Lin X, Xu J, Sun Y, Li E *et al.* Performing optical logic operations by a diffractive neural network. *Light: Science & Applications* 2020; **9**. doi:10.1038/s41377-020-0303-2.

25 Lin X, Rivenson Y, Yardimci NT, Veli M, Luo Y, Jarrahi M *et al.* All-optical machine learning using diffractive deep neural networks. *Science* 2018; **361**: 1004–1008.

26 Li J, Mengu D, Luo Y, Rivenson Y, Ozcan A. Class-specific differential detection in diffractive optical neural networks improves inference accuracy. *Adv Photon* 2019; **1**: 1.

27  Mengu D, Luo Y, Rivenson Y, Ozcan A. Analysis of Diffractive Optical Neural Networks and Their Integration With Electronic Neural Networks. *IEEE Journal of Selected Topics in Quantum Electronics* 2020; **26**: 1–14.

28  Li J, Mengu D, Yardimci NT, Luo Y, Li X, Veli M *et al.* Spectrally encoded single-pixel machine vision using diffractive networks. *Science Advances* 2021; **7**: eabd7690.

29  Kulce O, Mengu D, Rivenson Y, Ozcan A. All-optical information-processing capacity of diffractive surfaces. *Light Sci Appl* 2021; **10**: 25.

30  Kulce O, Mengu D, Rivenson Y, Ozcan A. All-optical synthesis of an arbitrary linear transformation using diffractive surfaces. *Light Sci Appl* 2021; **10**: 196.

31  Mengu D, Zhao Y, Yardimci NT, Rivenson Y, Jarrahi M, Ozcan A. Misalignment resilient diffractive optical networks. *Nanophotonics* 2020; **0**. doi:10.1515/nanoph-2020-0291.

32  Mengu D, Rivenson Y, Ozcan A. Scale-, Shift-, and Rotation-Invariant Diffractive Optical Networks. *ACS Photonics* 2021; **8**: 324–334.

33  Luo Y, Zhao Y, Li J, Cetintas E, Rivenson Y, Jarrahi M *et al.* Computational Imaging Without a Computer: Seeing Through Random Diffusers at the Speed of Light. *arXiv:210706586 [physics]* 2021.http://arxiv.org/abs/2107.06586 (accessed 22 Jul2021).

34  Rahman MSS, Ozcan A. Computer-free, all-optical reconstruction of holograms using diffractive networks. *arXiv:210708177 [physics]* 2021.http://arxiv.org/abs/2107.08177 (accessed 7 Oct2021).

35  Luo Y, Mengu D, Yardimci NT, Rivenson Y, Veli M, Jarrahi M *et al.* Design of task-specific optical systems using broadband diffractive neural networks. *Light: Science & Applications* 2019; **8**: 1–14.

36  Veli M, Mengu D, Yardimci NT, Luo Y, Li J, Rivenson Y *et al.* Terahertz pulse shaping using diffractive surfaces. *Nature Communications* 2021; **12**: 37.

37  Kingma DP, Ba J. Adam: A Method for Stochastic Optimization. *arXiv:14126980 [cs]* 2014.http://arxiv.org/abs/1412.6980 (accessed 16 Jun2018).

# Figures

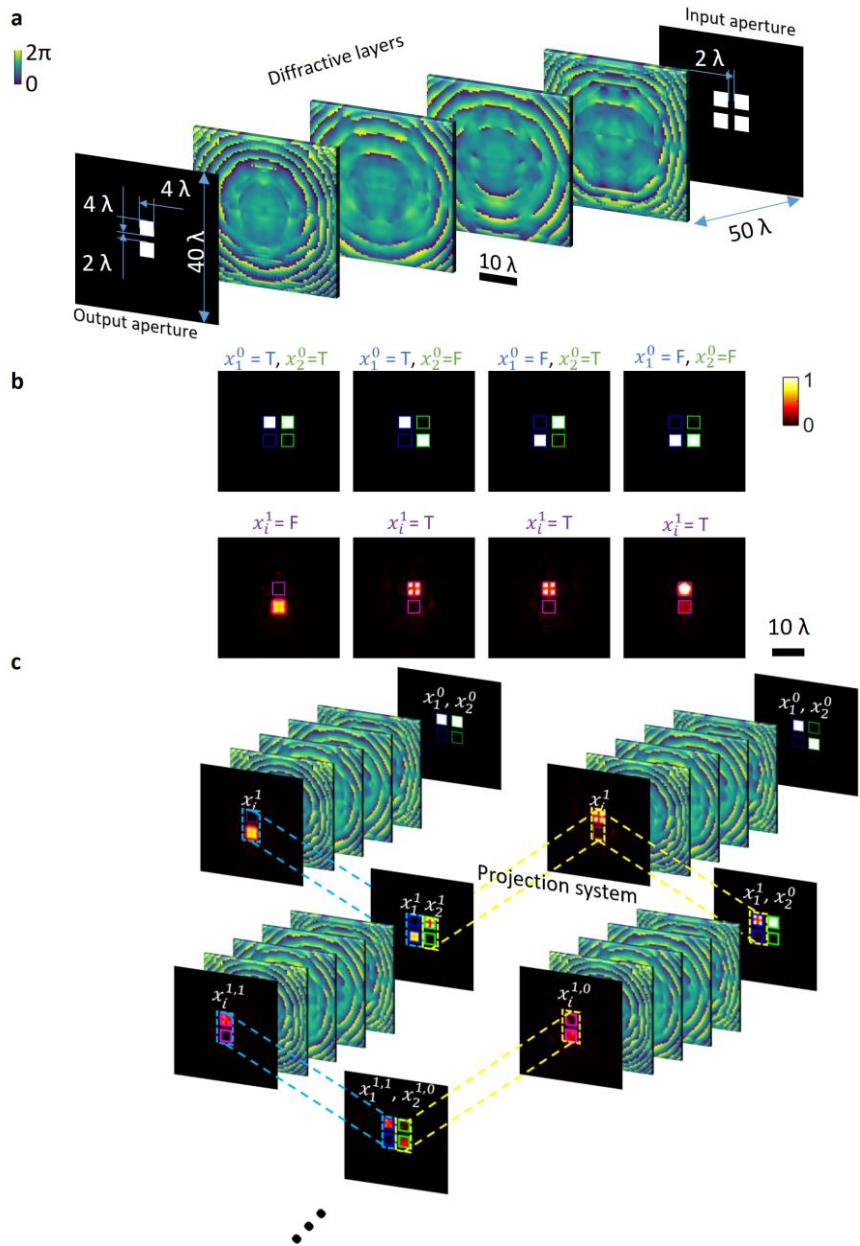

**Figure 1. Schematic of cascadable all-optical NAND gates using diffractive networks. a** Design of a cascadable diffractive NAND gate. **b** The input plane intensity profile of the diffractive NAND gate with all the possible combinations of ideal optical logical values, i.e., $(x_1^0, x_2^0)$ = (T,T), (T,F), (F,T) and (F,F), as well as the corresponding optical intensity profiles at the output plane of the diffractive network, with output optical logical values of $x_i^1$=(F, T, T, T) respectively. **c** Schematic of cascading diffractive NAND gates.

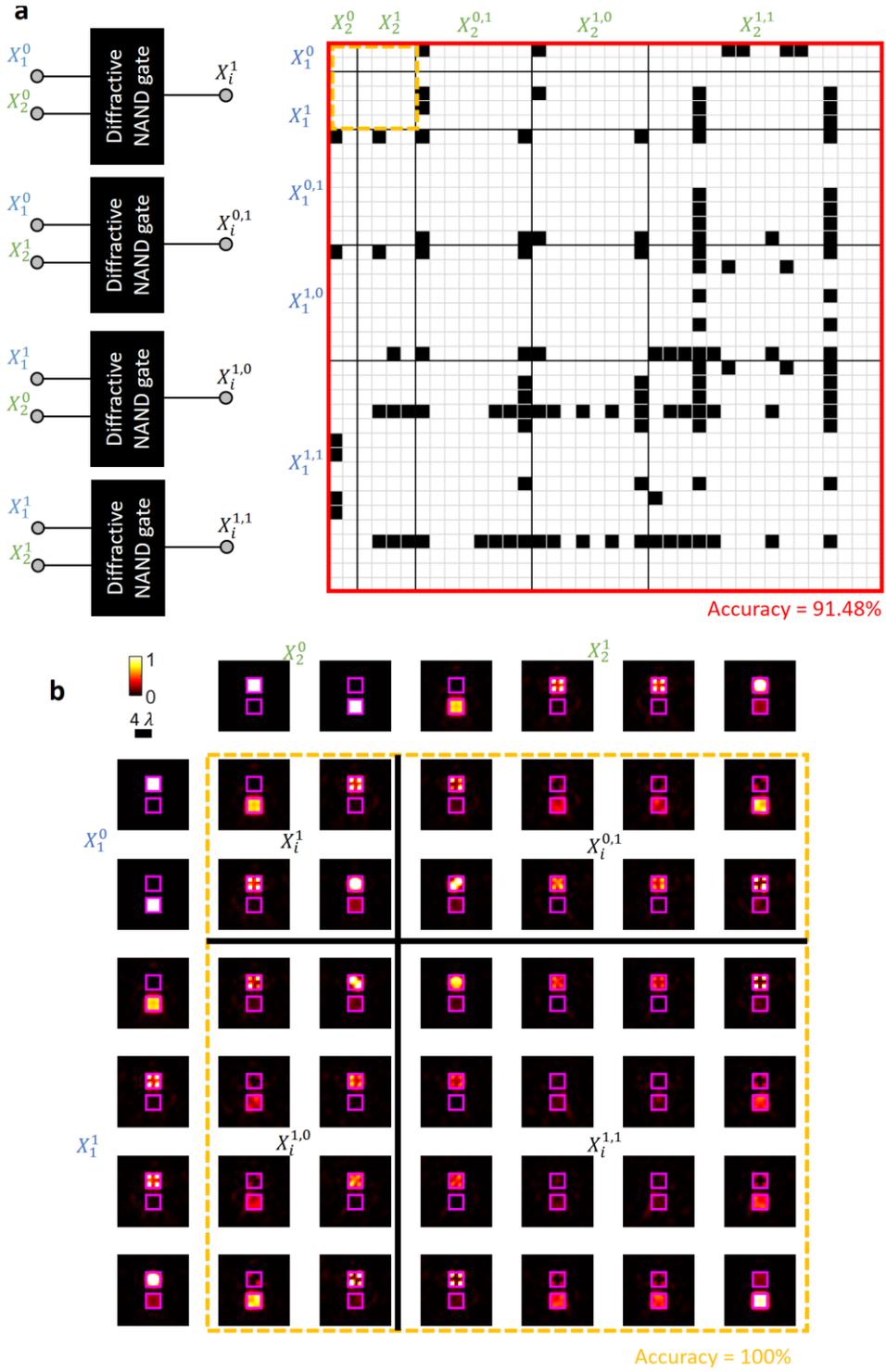

**Figure 2. Performance of cascaded diffractive NAND gates**. **a** A design map for cascading diffractive NAND gates. The black squares indicate the input wave combinations that lead to a miscalculation, i.e., an inference error. **b** The output plane intensity profiles of the optical waves that belong to sets $x_i^1$, $x_i^{1,0}$, $x_i^{0,1}$ and $x_i^{1,1}$ as well as their corresponding input waves.

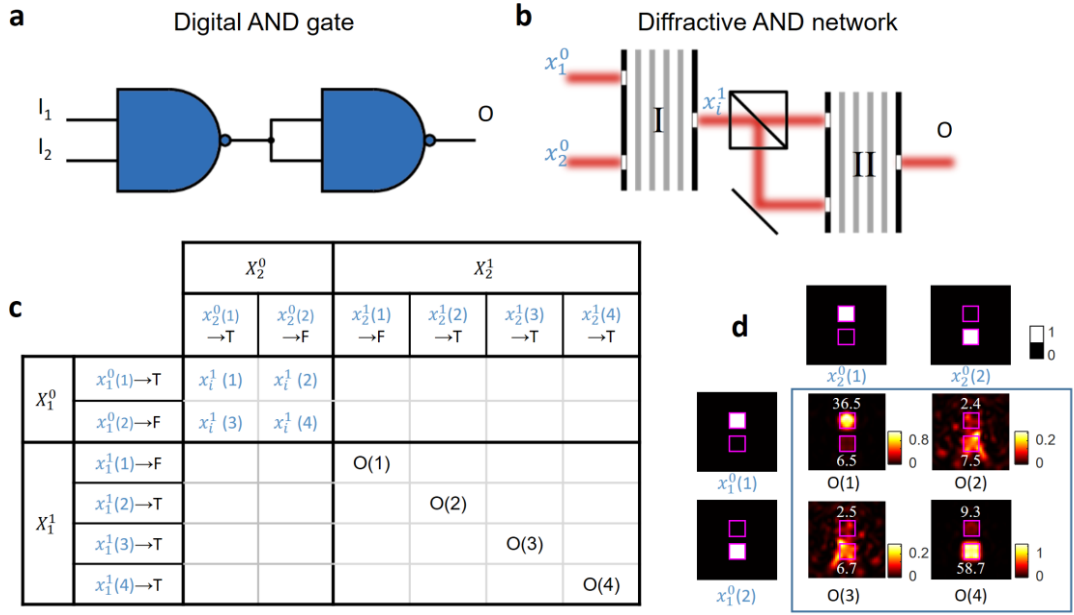

**Figure 3. Logical AND operator that is composed of cascaded diffractive NAND gates. a** Digital implementation of an AND operator using NAND gates. **b** All-optical AND gate design that is composed of cascaded diffractive NAND gates. **c** A portion of the design map (Fig. 2(b)) showing the correctness of the intermediate all-optical calculation steps ($x_i^1$) and the output of the AND gate (O) under different input combinations. **d** The intensity profiles at the output plane of the final diffractive NAND gate. The inserted numbers in white font color indicate the relative optical signal within each aperture.

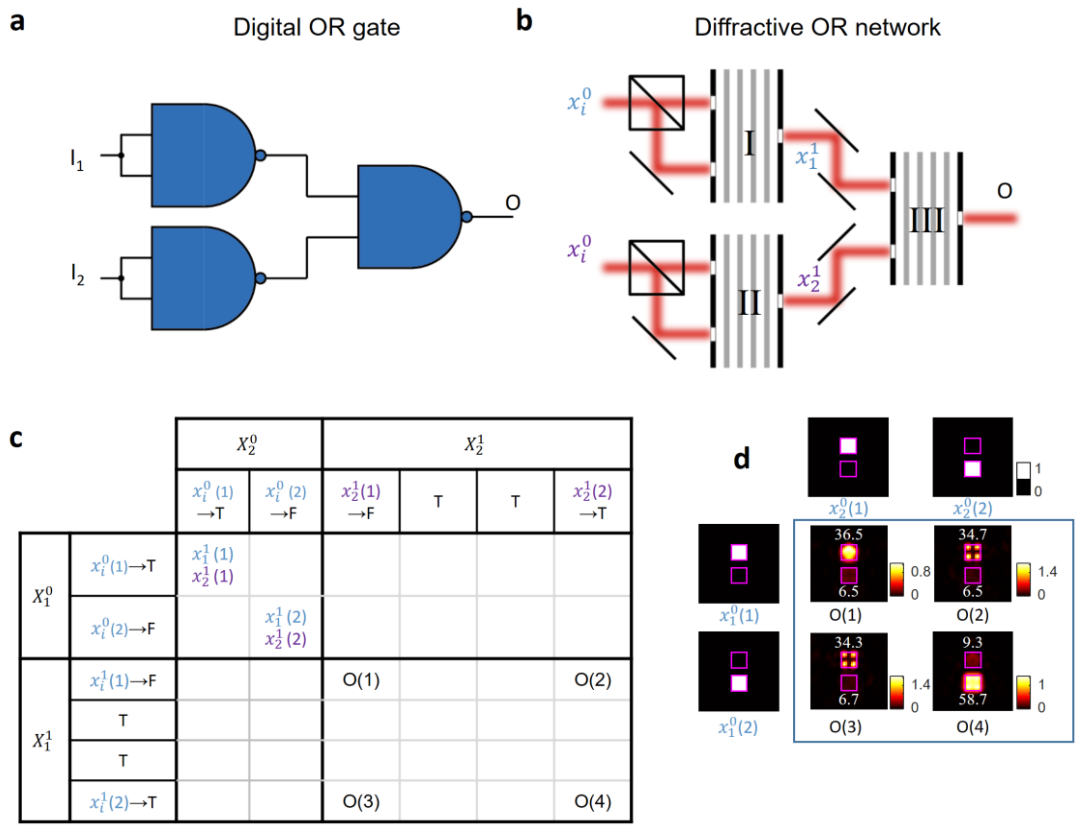

**Figure 4. Logical OR operator that is composed of cascaded diffractive NAND gates. a** Digital implementation of an OR operator using NAND gates. **b** All-optical OR gate that is composed of cascaded diffractive NAND gates. **c** A portion of the design map (Fig. 2(b)) showing the correctness of the intermediate all-optical calculation steps ($x_i^1$) and the output of the OR gate (O) under different input combinations. **d** The intensity profiles at the output plane of the final diffractive NAND gate. The inserted numbers in white font color indicate the relative optical signal within each aperture.

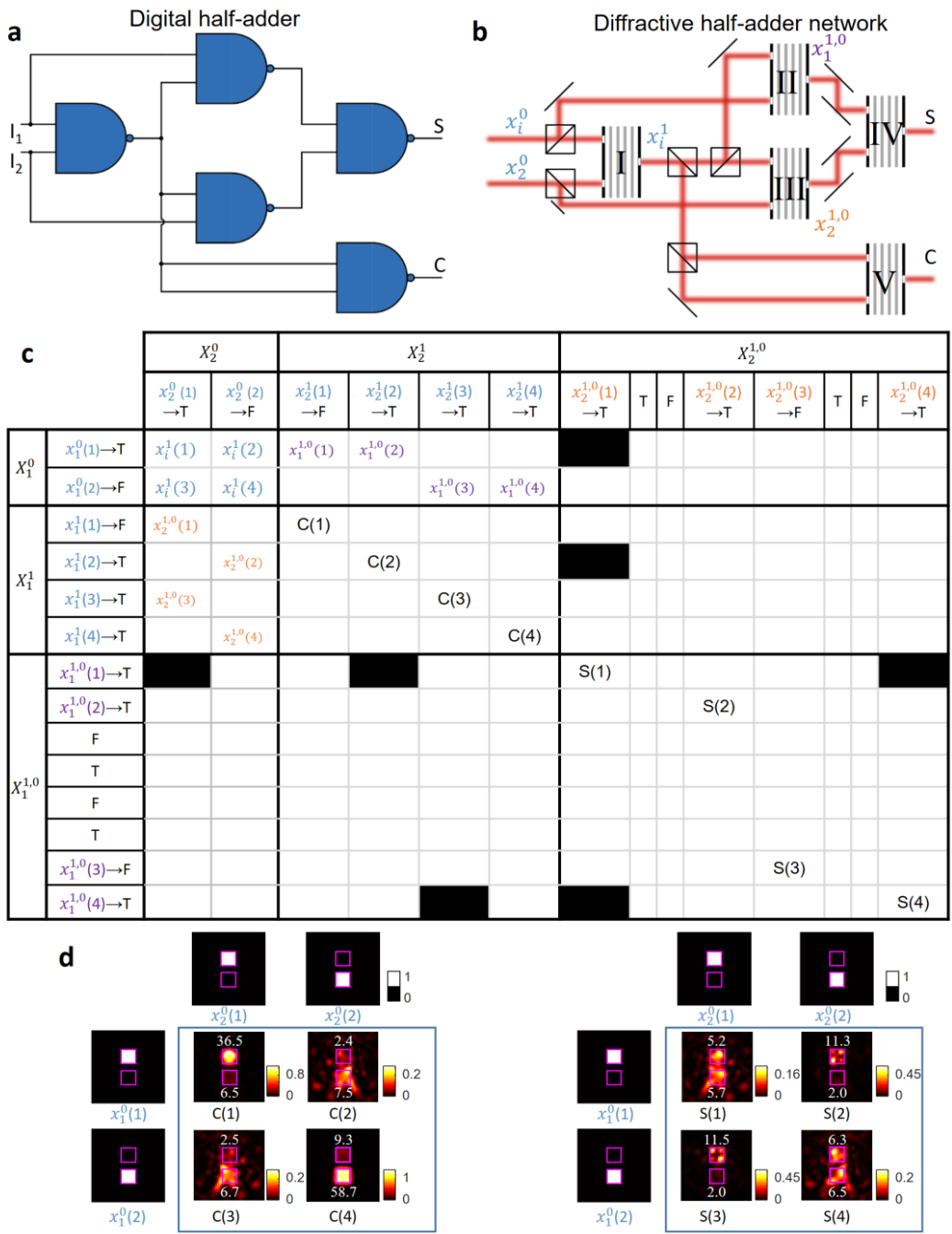

**Figure 5. A half-adder that is composed of cascaded diffractive NAND gates. a** Digital implementation of a half-adder using NAND gates. **b** An all-optical half-adder composed of cascaded diffractive NAND gates. **c** The correctness of the intermediate all-optical calculation steps ($x_i^1$) and the sum (S) and carry (C) of the half-adder using different input combinations. **d** The intensity profiles at the output plane of the diffractive NAND IV (representing the sum) and the diffractive NAND V (representing the carry) gates. The inserted numbers in white font color indicate the relative optical signal within each aperture.

24